\documentclass{elsart}
\usepackage{amssymb}
\usepackage{amsmath}
\usepackage{graphics}
\newcommand{\bmath}[1]{\mbox{\boldmath ${#1}$}}
\newcommand{\dd}{\mbox{\rm d}}
\begin{document}
\begin{frontmatter}

\title{\large{\bf Singlet-to-triplet ratio in the deuteron
breakup reaction  $\bmath{pd\to pnp}$ at $\bmath{585}$ MeV}}
\author[dubna,almata]{Yu.N.~Uzikov}\footnote
{Corresponding author:
Yu.N.~Uzikov, 
 Laboratory of Nuclear Problems, JINR,
Dubna, 141980, Moscow Region, Russia;\\
e-mail address: uzikov@nusun.jinr.ru;\\
FAX: 7 09621 66666},
\author[dubna]{V.I.~Komarov},
\address[dubna]{JINR, LNP, Dubna, 141980, Moscow Region, Russia}
\address[almata]{ Kazakh State University, Almaty,  480121 Kazakhstan}
\author[juelich]{ F.~Rathmann},
\author[juelich]{ H.~Seyfarth}
\address[juelich]{Institut f\"ur Kernphysik,
 Forschungszentrum J\"ulich, 52425 J\"ulich, Germany}

\begin{abstract}
 Available experimental data on the exclusive  $pd\to pnp$ reaction
 at 585~MeV show a narrow peak
 in the  proton-neutron final-state interaction region.
 It was supposed previously,
 on the basis of a phenomenological analysis of the shape of this peak,
 that the final  spin-singlet
 $pn$ state provided about one third of the observed cross section.
 By comparing 
 the absolute value of the measured cross section
 with that of $pd$ elastic scattering  using the
  F\"aldt-Wilkin extrapolation theorem,
 it is shown here  that the
  $pd\to pnp$  data  can be explained mainly by the spin-triplet
 final state with a singlet admixture  of a few percent.
 The smallness of the  singlet  contribution is compatible with
 existing  $pN\to pN\pi$ data 
and  the one-pion exchange  mechanism  of the $pd\to pnp$  reaction.

\end{abstract}
\vspace{5mm}
\vspace{8mm}
\noindent
\begin{keyword}
 proton deuteron  scattering,
 final state interaction
\begin{PACS}
 13.75.Cs, 25.45.D, 25.10.+s\\[1ex]
\end{PACS}
\end{keyword}
\end{frontmatter}

\newpage
\baselineskip 4ex

 Recently, the $NN\to NN\pi$  reactions with the formation of a
 spin-singlet $NN$ pair in the final state
 have received    a renewed   interest.
 Analyzes of the experi-\\mental data obtained at COSY \cite{anke},
CESLIUS \cite{celsius} and LAMPF \cite{HG},
 employing the largely model-independent approach 
 of Ref. \cite{FW1}, show that
 the singlet channel is strongly suppressed in the $pp\to pn\pi^+$
 reaction at proton kinetic energies between 300 and 800 MeV
\cite {FW3,bfw,uzcw}.
  Direct measurements
 of the singlet channel in the reaction $pp\to pp\pi^0$  at
 RCNP \cite{maeda} and  CELSIUS
 \cite{bilger} at $300 - 400$ MeV indicate  a singlet-to-triplet
($s/t$) ratio  of about 1\% 
 in collinear kinematics, 
 which  increases  up to $\sim 10$\%  as the cm scattering angle 
 approaches $90^\circ$.
 The  dominance of the triplet state
 can be related to the  excitation of a $\Delta$-isobar
 in the intermediate state \cite{uzcw}.

 The measured pion production cross section in $pp$ collision
 allows one to  estimate qualitatively  the $s/t$
 ratio in the deuteron breakup 
 reaction $pd\to \{pn\}p$, when  the quasi-bound
 $\{pn\}$ pair is observed in the
 final state interaction  (fsi) region and the second 
 proton is detected 
 at large cm scattering  angle ($\theta^*>90^\circ$).
 It is well known  that in backward elastic $pd$ scattering
 $pd\to dp$ the triangle diagram of one-pion exchange
 with the subprocess $pp\to d\pi^+$ considerably
 contributes in the $\Delta$-region \cite {kolybasm}.  This
 mechanism   describes  well the energy dependence of the
 $pd\to dp$ cross section at $\theta ^*=180^\circ$ and,  in addition,
 explains the
 qualitative  agreement between the proton vector analyzing power $A_y$
 from  $pp\to d\pi^+$ and $pd\to dp$,
 observed  in the $\Delta$-region \cite {biegert}.
 If one assumes that the triangle diagram with one-pion exchange 
 dominates   in the break-up $pd\to \{pn\}p$  at
 large  scattering angles, one would expect in this reaction
  a similar $s/t$ ratio of a few percent, as observed in $pp\to pn\pi^+$.
 For the $\Delta$ mechanism of the $pd\to pnp$ reaction, which dominates
 the one-pion exchange triangle diagram, the product
 of  spin and isospin factors yields a  $s/t$ ratio of $\frac{1}{27}$
 \cite{uzi2000}.
 In contrast, one should expect a  higher $s/t$ ratio
 of about $\frac{1}{3}$  for 
 the  one-nucleon exchange mechanism of the deuteron breakup \cite{uzi2000}. 
 It was suggested in Refs. \cite{uzi2000,imuz90,smuz98}
 to  directly measure  the singlet channel in the reaction
 $pd\to (pp)(0^o)+n(180^o)$  with a 
 {\it pp} pair of low relative energy $E_{pp}=~0-5$ MeV emitted in forward 
 direction  and a neutron  going backward. Due to  a considerable
 suppression of the $\Delta $-mechanism  in this reaction \cite{uzi2000}
 other mechanisms, more sensitive to the short-range structure of the
 deuteron, are expected to become  important \cite{cosy20}.

 Recent  experimental data   on
 the deuteron breakup reaction $dp\to pnp$ with   two outgoing
 nucleons in the fsi region were obtained  at Saclay \cite{bfw} at
 $T_d=1.6$ GeV in semi-inclusive
 kinematics  and  at Dubna \cite{dubna98} at $T_d=$~2-5 GeV.
 Earlier, a kinematically complete exclusive
 experiment  had been  performed at  Space Radiation Effects Laboratory
 (SREL) in Virginia \cite{witten} at a proton beam  kinetic
 energy of  $T_p=585$~MeV, 
 covering  a region of low relative neutron-proton energy
  $E_{np}=0-5$ MeV outside  of quasi-free $pN$-kinematic.
 A clear peak was observed in the five-fold cross
 section at $E_{np}\sim 0$. Using  the Migdal-Watson
 approximation \cite{Watson,GW},
 the authors  of Ref. \cite{witten}  described the shape of the fsi peak
 by  assuming a $s/t$ ratio of one third, which  corresponds
 to the spin statistical weights of the singlet and triplet states.
 A smaller  $s/t$ ratio of  about $ 10$\% was obtained from 
 the  data of Ref. \cite{bfw}.  The difference  is possibly  related
 to the different cm scattering angles of protons
  ($\theta^* \sim 90^\circ$ in Ref. \cite{witten}  and 
 $\theta^* \sim 180^\circ$ in Ref. \cite{bfw} ).

 However, the fitting procedure described  in Ref. \cite{witten}
 is rather  ambiguous since the absolute value of neither the triplet
 nor the singlet
 cross section is known and was  arbitrarily introduced.
 The $s/t$ ratio can be deduced in principle  from the data,
 taking into account only
 the strong difference in shape of the singlet and triplet
 peaks (see, for example, Ref. \cite{anke}).  Unfortunately,
the low resolution in $E_{np}$ and   limited
 statistics in the peak  do not allow one
 to effectively use this procedure for the  data of Ref. \cite{witten}.
 In this case  the knowledge of the
 absolute value of the triplet (or singlet) cross section is necessary
 in order to determine the $s/t$ ratio.
 The triplet cross section can be calculated in a model-independent
 way  in terms of the large angle 
 proton-deuteron elastic scattering.
 Here we employ the approach described in  Refs. \cite {FW1,FW3,bfw,uzcw}
 to determine the triplet cross section and on this basis
 reanalyze the data of Ref. \cite{witten}.

The SREL data are shown in Fig.~1
as a function of the detected proton momentum.
At energies  $E_{np}$ of about 1~MeV  the
cross section is strongly influenced by the $np$ fsi. The shape of this
peak is well described by the Migdal-Watson 
formulae~\cite{Watson,GW}, which take into account the nearby poles
in the fsi triplet ({\it t}) and singlet ({\it s}) $pn-$scattering
 amplitudes
\begin{equation}
\label{mwatson}
 d\sigma_{s(t)} =FSI_{s(t)}(k)\, K\, |A_{s(t)}|^2.
\end{equation}
 Here $A_{s(t)}$ is the production matrix element for the
 singlet (triplet) state, $K$ is
 the kinematical factor, and $FSI_{s(t)}$ is the  Goldberger-Watson
 factor \cite{GW}. The latter   can be written in the  form
\begin{equation}
\label{gwfsi}
 FSI_{i}=\frac{k^2+\beta_i^2}{k^2+\alpha_i^2},
\end{equation}
 where $i=s,t$.
 The relative momentum in the $pn$ system at
 the relative kinetic energy $E_{np}=k^2/m_N$ is denoted by $k$,
 $m_N$ is the nucleon mass.
The parameters $\alpha$ and $\beta$  are determined by known 
 properties of the on-shell $NN$-scattering 
 amplitudes at low energies: 
  $\alpha_t=0.232$ fm$^{-1}$, $\alpha_s=-0.04$ fm$^{-1}$,
 $\beta_t=0.91$ fm$^{-1}$, $\beta_s=0.79$ fm$^{-1}$ \cite{machl}.
  Important new information
 on the  mechanism of  $pd\to pnp$ and off-shell
 properties of the {\it NN} system is hidden in the  matrix elements
 $A_{s(t)}$, in particular in the ratio
\begin{equation}
\label{zeta}
\zeta=\frac{|A_{s}|^2 }{|A_{t}|^2 }.
\end{equation}

One can find  from Eqs. (\ref{mwatson}) and  (\ref{zeta})
the following parametrization
 for the full singlet plus triplet cross section \cite{uzcw}
\begin{equation}
\label{singlet}
{\dd\sigma_{s+t}}
=\left (1+\zeta\,\frac{FSI_s}{FSI_t}\right )
{\dd\sigma_t},
\end{equation}
 where $\dd\sigma_t$ is the triplet cross section.
 The second term in the brackets of Eq.~(\ref{singlet})
 corresponds to the singlet contribution.

 Using the
 F\"aldt-Wilkin extrapolation \cite{FW1}, which relates the  bound
 and  the scattering S-wave functions in the triplet state
  at  short {\it pn} distances
 $r<1$~{\rm fm}, and  by taking into account the short-range character of the
 interaction mechanism, one can find a definite  relation between
 the matrix elements of the  $pd\to \{{pn}\}_t\,p$ and
 $pd\to dp$ reactions \cite{FW1,bfw}.
 The triplet differential cross section in the laboratory system
 can then be written  as
\begin{equation}
\label{yuri}
\frac{\dd^5\sigma_t(pd\to pnp)}{\dd p_1\,\dd\Omega_1\,\dd\Omega_2}
= \frac{1}{16 \pi^3}\, \frac{p_1^2 p_2^3\, s\,f^2(k^2)   }{ p_0\, m_d\, E_1\,
|p_2^2\, E_n -{\bf p}_2\cdot {\bf p}_n E_2|}
\frac{\dd\sigma}{\dd\Omega^*}(pd\to dp),
\end{equation}
 where 
\begin{equation}
\label{fsi}
f^2(k^2) = \frac{2\pi\,m_N}{\alpha_t(k^2+\alpha_t^2)}
\end{equation}
 is the F\"aldt-Wilkin factor \cite{bfw},
 $\dd\sigma/ \dd\Omega^*$ is the $pd\to pd$
 cm cross section. In Eq. (\ref{yuri}) $s$ denotes 
 the squared invariant mass of the $pd$ system, $m_d$ is the deuteron mass,
 $p_0$ is the beam momentum, $E_i$ and ${\bf p}_i$ ($i=1,2,n$) are the
 laboratory energy and momentum of the $i$-th nucleon in the final state.
 The indices  1 and 2   refer to the  protons and
 the neutron is referred as {\it n}.
 The proton scattering  angles in the $pd\to \{pn\}p$ and  $pd\to dp$
 processes can be related to each other, if
 the difference between the effective mass of
 the final $\{pn\}$ system and that of the deuteron is disregarded,
 as suggested in Ref. \cite{uzcw}.
 The result presented by Eq.~(\ref{yuri}) should i) be valid at low relative
 energies $E_{np}$, ii) be independent of the form of the $NN$-potential and
 details of the large-angle $pd$-scattering mechanism, and iii) it
 automatically includes the fsi effects in the
 triplet $pn$ system. On the other hand, this method cannot be used for
 small-angle pd-scattering since the $NN$-scattering and bound-state
 wave functions are very different at large $NN$ distances, i.e.  at low
 transferred momenta.

 The  value of the  differential cross section $\dd\sigma/ \dd\Omega^*$
  in Eq. (\ref{yuri}) at $T_p=590$ MeV and
 $\theta^*=92.7^\circ$ amounts to
 $\left ( 30.4{ \pm 0.8(stat.) \atop \pm 2.9(syst.)}\right ) \mu $b/sr
 \cite{alder}. 
 The SREL experiment \cite{witten}
 was carried out at almost the same scattering angle
 ($\theta_2^*=93.95^\circ$ for  $E_{np}=0$).
 Other available data \cite{bosch,albr} give larger values
 for the $pd\to dp$ cross section under similar kinematic conditions.
 Therefore,
 in order to estimate an {\it upper} limit for the $s/t$ ratio we use here
 only the data from Ref. \cite{alder}. 
 As one can see from Fig. 1a,
 the triplet cross section calculated using
 Eq.~(\ref{yuri}) (dashed line)
 overshoots the experimental points in the central region around  
  $E_{np}\sim 0$, but agrees with   the data
 for  $E_{pn}> 3$ MeV. However,
 a sizable effect arises from averaging 
 of the theoretical results
over the experimental
 angular acceptance and resolution of the spectrometer. In order to
  take these into account, we have carried out a
 five-dimensional integration
 of the cross section from  Eq. (\ref{yuri}) with  Gaussian distributions,
 where  smearing parameters $\sigma_\theta=2.55^\circ$ for the polar
 angles of   and $\sigma_p/p=0.015 $
 for  the momentum $p$  were used in accordance
 with Ref.~\cite{witten}. For the azimutal angles
 $\phi_{1}$ and $\phi_{2}$  the averaging was carried out
 in the interval $\Delta \phi=\pm 0.4^\circ$ with a rectangular distribution.

  After smearing 
 we obtain good agreement both in the shape and in absolute value
 between  one data set  (Fig. 1a)  and  a pure
 triplet contribution of the final $pn$ pair with a
  $\chi^2= 0.7$.
 A small singlet contribution,
 corresponding to  $\zeta=0.02$,
 does not contradict  the data ($\chi^2=0.9$),
 whereas  larger values 
 $\zeta=0.05$ ($\chi^2=1.8$) and  $\zeta=0.10$ ($\chi^2=4.6$)
 result in too large a cross section in the vicinity of $E_{np}=0$.
 The other data set  (Fig. 1b),  obtained at a different
 magnetic field  setting, shows  also dominance of the triplet contribution
  and allows a small singlet fraction:
 ($\zeta,\chi^2$)= (0.0, 2.4,), (0.02, 2.1), (0.05, 2.3), (0.10, 4.0).
  However, in this case   the $\chi^2$ becomes worse. 
 Under assumption of $\zeta=\frac{1}{3}$, made in Ref. \cite{witten},
 the absolute value of the  cross section in the region around $E_{np}=0$
 results by a factor 2.5 - 3 too high compared with the data.

 The accuracy of the  approximation by  Eq.~(\ref{yuri}) is  estimated 
 in Refs. \cite{FW1,FW3,bfw,uzcw} and \cite{smuz98}
 to be better than 5\%  for $E_{np}\leq 3$~MeV.
 This error  arises  from  variations  of
 the bound and scattering $NN$ wave functions at short distances
 for low $E_{np}$. The error of the $pd\to dp$ input is
 $\approx 9$\%~\cite{alder}.
 The systematic uncertainties 
 in the measured  $d\sigma_{s+t}$ are not given in \cite{witten}, here
 we assume them   not to  exceed 10\%. Combining all uncertainties
 given above,  the  $d\sigma_{s+t}$
 in Eq. (\ref{singlet}) is uncertain within  15\%. If the measured
 cross section given in Fig.~1a is scaled by factors ranging 
  from 0.85 to 1.15, our $ \chi^2(\zeta)$ analysis  shows 
 that  the resulting $\zeta$'s for minimum $ \chi^2$  range  
 from  $ +0.035$ to $-0.030$  with the corresponding uncertainties 
 $\Delta \zeta$ ranging from ${+0.065 \atop -0.055 }$ to
 ${+0.040 \atop -0.035 }$, respectively. 
 This implies  that  $\zeta$  and $\Delta \zeta$ are both of the order
  of a few percent, and thus are  substantially smaller
  than the spin-statistical factor  of $\frac{1}{3}$ assumed
 in Ref. \cite{witten}.

  The  matrix element squared 
 $|M|^2$ shown in Fig. 3 was  obtained  in  Ref. \cite{witten}
 by dividing the raw data   point by point  by  a Monte Carlo
  $E_{np}$ energy distribution, that includes the
 phase space factor. By this procedure the  authors of 
 Ref. ~\cite{witten} minimized the effects from averaging over the
 detector acceptance.
 In contrast to the production  matrix element 
 $|A|$, defined  by Eq.~(\ref{mwatson}),
 the complete matrix element $|M|$ contains the fsi.
 The authors of Ref. \cite{witten}
 found that the spin-statistical fraction
 of the singlet of $\frac{1}{3}$ describes the measured
 data. However, the experimental data   contain
 considerable uncertainties. Therefore, according to our calculations,
 they do not constrain the singlet fraction  strongly enough.
 As can be seen  from Fig.~3,  values
 $\zeta =0.05$ and  $0.30$ allow one  to fit the experimental data
  equally well ($\chi^2= 1.4$ and $\chi^2= 0.9$, respectively ),
 if the absolute value
 of the matrix element $|M|^2$,  not given in Ref. \cite{witten},
 is  treated as a free parameter.
 The small value of $\zeta$,
 which we found from the 
 cross section, is compatible  
 with the value $\zeta =0.19 { +0.32 \atop -0.16}$,
 resulting from our  analysis of  the $ \chi^2(\zeta)$ 
 distribution for  the $|M|^2$ data.

 To improve the sensitivity to the $s/t$ ratio using the
 extrapolation theorem of Ref. \cite{FW1,bfw}, 
  the ratio of the $pd\to pnp$ and
 $pd\to dp$ cross sections has to be established better by a
 measurement of both reactions in the same experiment.
 A new measurement
 of the ${\vec p}{\vec d}\to pnp$ reaction at
 the ANKE spectrometer of the
 proton synchrotron COSY-J\"ulich will put more stringent 
 limits on the $s/t$  ratio by detecting   both protons
  in the forward-forward or  forward-backward  directions at beam energies
 $T_p=0.5-2.5$~GeV~\cite{cosy20}.

 In conclusion, by comparing  the $pd\to pnp$ cross section
  at 585~MeV with that of $pd\to dp$ on the basis of scattering theory,
 we found that the final state spin-triplet contribution is dominant
 allowing  a singlet  contribution of a few percent. 
 This  result  is in agreement with  existing experimental
 data on the $s/t$ ratio in the reaction
 $pN\to pN\pi$ and supports  the dominance of the
 triangle diagram with the subprocesses $pN\to {pN}\,\pi$
  in the reaction $pd\to pnp$.
 
 The authors would like to thank   C. Wilkin  
  for helpful remarks.  Two of us (Yu. U. and V. K.)
 gratefully acknowledge financial support  and warm hospitality
 at  IKP of   Forschungszentrum J\"ulich.
 This work was supported in part by  the Heisenberg-Landau program and
  BMBF  grant KAZ 99/001.

\input epsf
\begin{figure}[t]
\begin{center}
\vspace*{-1cm}
\mbox{\epsfxsize=7.5in \epsfbox{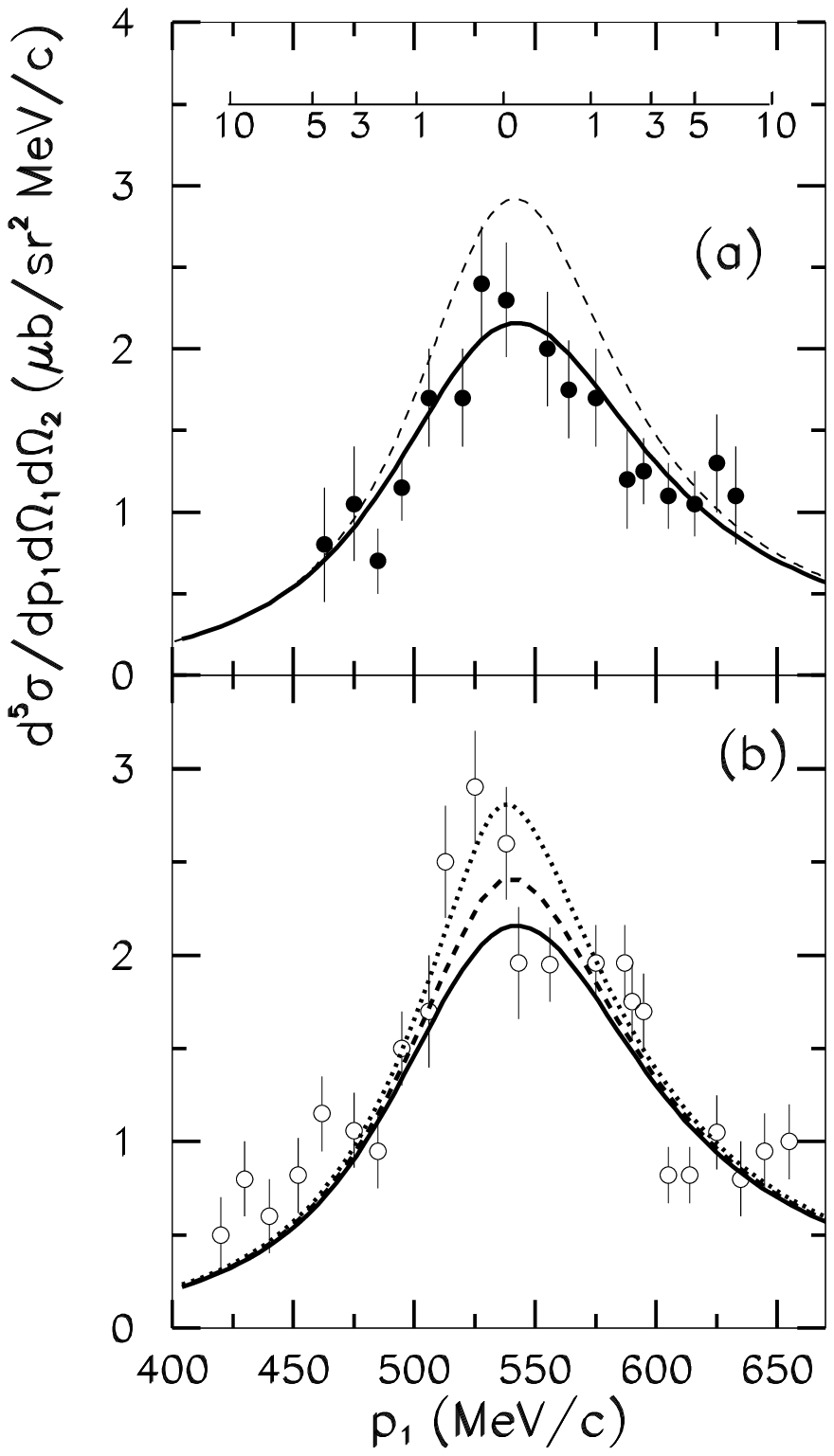}}
\caption{  Experimental cross section  (points) of
 the $pd\to pnp$ reaction from Ref. \protect\cite{witten}
  at beam   energy $585$ MeV and proton  laboratory scattering angles
 $\theta_1=41^o$, $\theta_2=61^o$  as function
 of the proton momentum 
  in comparison with our calculations.
 {\bf a}) The pure triplet contribution
  calculated  with corrections taking into account
  the experimental resolution (full line)
 and   without (dashed), as explained in the text.
 The upper scale shows the relative energy (in MeV)
 of the $pn$-pair for $\theta_1=41^o$.
 {\bf b}) The same observable  as in  a) but  for another magnetic
     field setting,  compared with calculations including  the corrections
     for different   $s/t$ ratios   
     $\zeta=0.0$ (full line), $0.02$ (dashed), and  $\zeta=0.05$
    (dotted).
}
\end{center}
\label{fig1}
\end{figure}

\eject
\begin{figure}[t]
\begin{center}
\vspace*{-4cm}
\mbox{\epsfxsize=6.5in \epsfbox{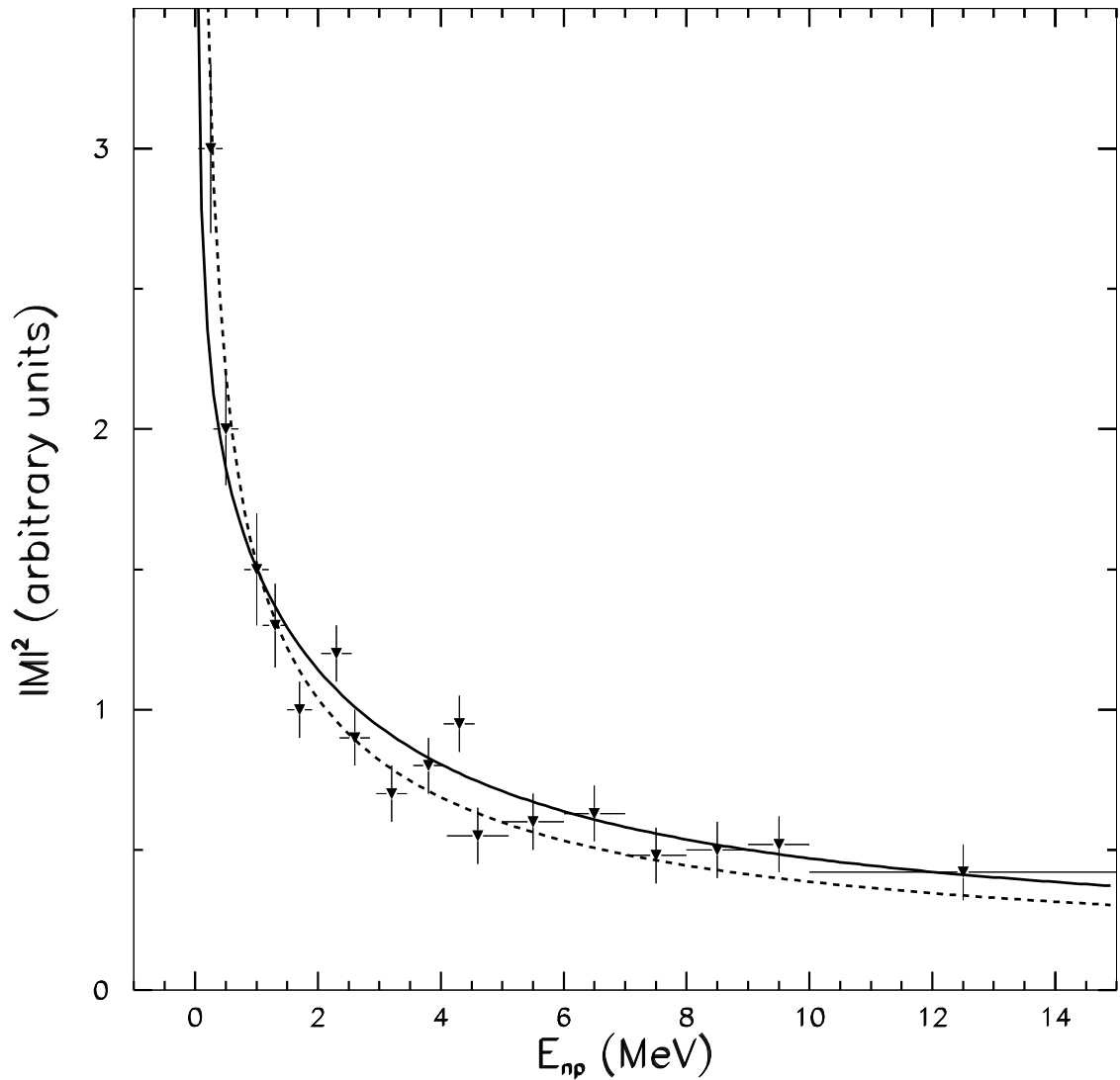}}
 \caption{ The squared matrix element, as obtained in
   Ref. \protect\cite{witten},
 for  arbitrary normalization  is  well described  by 
  $\zeta=0.05$, $\chi^2= 1.4$ (full line ) and 
  $\zeta=0.30$ $\chi^2= 0.9$ (dashed).}
\label{fig3}
\end{center}
\end{figure}
\end{document}